\documentclass[11pt]{article}
\usepackage{jinstpub} 
\usepackage{subfigure}
\usepackage{soul}
\usepackage{hyperref}

\title{Improvement on the Linearity Response of PandaX-4T with new Photomultiplier Tube Bases}

\author[a]{Lingyin Luo,}
\author[h]{Deqing Fang,}
\author[c]{Ke Han,}
\author[c]{Di Huang,}
\author[c]{Xiaofeng Shang,}
\author[d,e]{Anqing Wang,}
\author[h]{Qiuhong Wang,}
\author[c,f]{Shaobo Wang,}
\author[a]{Siguang Wang,}
\author[g]{Xiang Xiao,}
\author[b,1]{Binbin Yan,\note{Corresponding author.}}
\author[i]{and Xiyu Yan}

\affiliation[a]{School of Physics, State Key Laboratory of Nuclear Physics and Technology, Peking University, Beijing 100871, China}
\affiliation[b]{Tsung-Dao Lee Institute, Shanghai Jiao Tong University, Shanghai, 200240, China}
\affiliation[c]{School of Physics and Astronomy, Shanghai Jiao Tong University, Key Laboratory for Particle Astrophysics and Cosmology (MoE), Shanghai Key Laboratory for Particle Physics and Cosmology, Shanghai 200240, China}
\affiliation[c]{ INPAC and School of Physics and Astronomy, Shanghai Jiao Tong University, Shanghai Laboratory for Particle Physics and Cosmology,\\Shanghai 200240, China}
\affiliation[d]{Research Center for Particle Science and Technology, Institute of Frontier and Interdisciplinary Science, Shandong University, Qingdao 266237, Shandong, China}
\affiliation[e]{Key Laboratory of Particle Physics and Particle Irradiation of Ministry of Education, Shandong University, Qingdao 266237, Shandong, China}
\affiliation[f]{SJTU Paris Elite Institute of Technology, Shanghai Jiao Tong University, Shanghai, 200240, China}
\affiliation[g]{School of Physics, Sun Yat-Sen University, Guangzhou 510275, China}
\affiliation[h]{Key Laboratory of Nuclear Physics and Ion-beam Application (MOE), Institute of Modern Physics, Fudan University, Shanghai 200433, China}
\affiliation[i]{School of Physics and Astronomy, Sun Yat-Sen University, Zhuhai, 519082, China}

\emailAdd{yanbinbin@sjtu.edu.cn}

\abstract{
  With the expanding reach of physics, xenon-based detectors such as PandaX-4T in the China Jinping Underground Laboratory aim to cover an energy range from sub-keV to multi-MeV.
  A linear response of the photomultiplier tubes (PMTs) is required for both scintillation and electroluminescence signals.
  Through a dedicated bench test, we investigated the cause of the non-linear response in the Hamamatsu R11410-23 PMTs used in PandaX-4T.
  The saturation and suppression of the PMT waveform observed during the commissioning of PandaX-4T were caused by the high-voltage divider base.
  The bench test data validated the de-saturation algorithm used in the PandaX-4T data analysis.
  We also confirmed the improvement in linearity of a new PMT base design with three more low radioactivity capacitors at later dynodes, which will be used to upgrade the PMT readout system in PandaX-4T.
}  


\begin{document}
\maketitle
\flushbottom

\section{Introduction}
In recent years, the liquid xenon (LXe) time projection chamber (TPC) has become a leading experimental technique for rare event searches in underground laboratories~\cite{Aalbers:2022dzr,DARWIN:2016hyl}.
Traditional dark matter direct detection experiments, such as PandaX-4T~\cite{PandaX-4T:2021bab}, XENONnT~\cite{XENON:2020kmp}, and LZ~\cite{LZ:2021xov}, have transitioned to the multi-ton-scale stage and expanded the physics reach beyond dark matter.
LXe detectors also explore additional physics, including double beta decay of $^{136}$Xe~\cite{silin2022DBD,2022XenonDBD,LZ:2019qdm,PandaX-II:2019euf}, double electron capture of $^{124}$Xe~\cite{XENON:2019dti}, and solar $^8$B neutrino searches using coherently elastic nuclear scattering (CE$\nu$NS)~\cite{4TB8, XENON:2020gfr}.
The energy range of interest spans from sub-keV to a few MeV, necessitating a stringent energy response in terms of linearity and resolution for dark matter and neutrino physics.

PandaX-4T employs two arrays of Hamamatsu R11410-23 photomultiplier tubes (PMTs)~\cite{Baudis:2013xva} to capture the prompt scintillation signals and delayed electroluminescence signals.
The photon response of the PMTs determines the detector response.
The initial version of the custom-designed high voltage divider and readout circuit base for the PMTs (referred to as PMT base later) in PandaX-4T focused on dark matter detection in the keV energy range~\cite{Zheng:2020kfp}.
To address the possible radioactivity of capacitors, only three capacitors were used at the last stages of the dynodes.
However, this design choice resulted in PMT waveform distortion and a non-linear energy response for signals at the MeV level.
To mitigate this effect, a de-saturation algorithm was developed for double beta decay analysis~\cite{silin2022DBD,XENON:2020iwh}.

An improved version of the PMT base was designed and verified to significantly extend the linear photon response range, ensuring that future PandaX-4T detectors would not suffer from PMT waveform distortion for physics signals in the MeV range.
A dedicated bench test was set up to confirm the original PMT base's waveform distortion and the new base's extended linearity range.
The improved PMT bases have been installed in PandaX-4T and will be further tested with detector data.

Moreover, for multi-site events that deposit energy within a narrow region, the latter signals are likely suppressed by the first large signal. The suppression effect is difficult to correct due to the complicated detector response. We design a modern base with five de-saturation capacitors to achieve better detector linearity at the MeV scale. Such a design can meet the requirements of the PandaX-4T and will be installed in the upcoming detector update.

\section{PandaX-4T detector and linearity requirement of PMTs}\label{sec:4T requirement}

\begin{figure}[tbp]
  \centering
  \includegraphics[width=\textwidth]{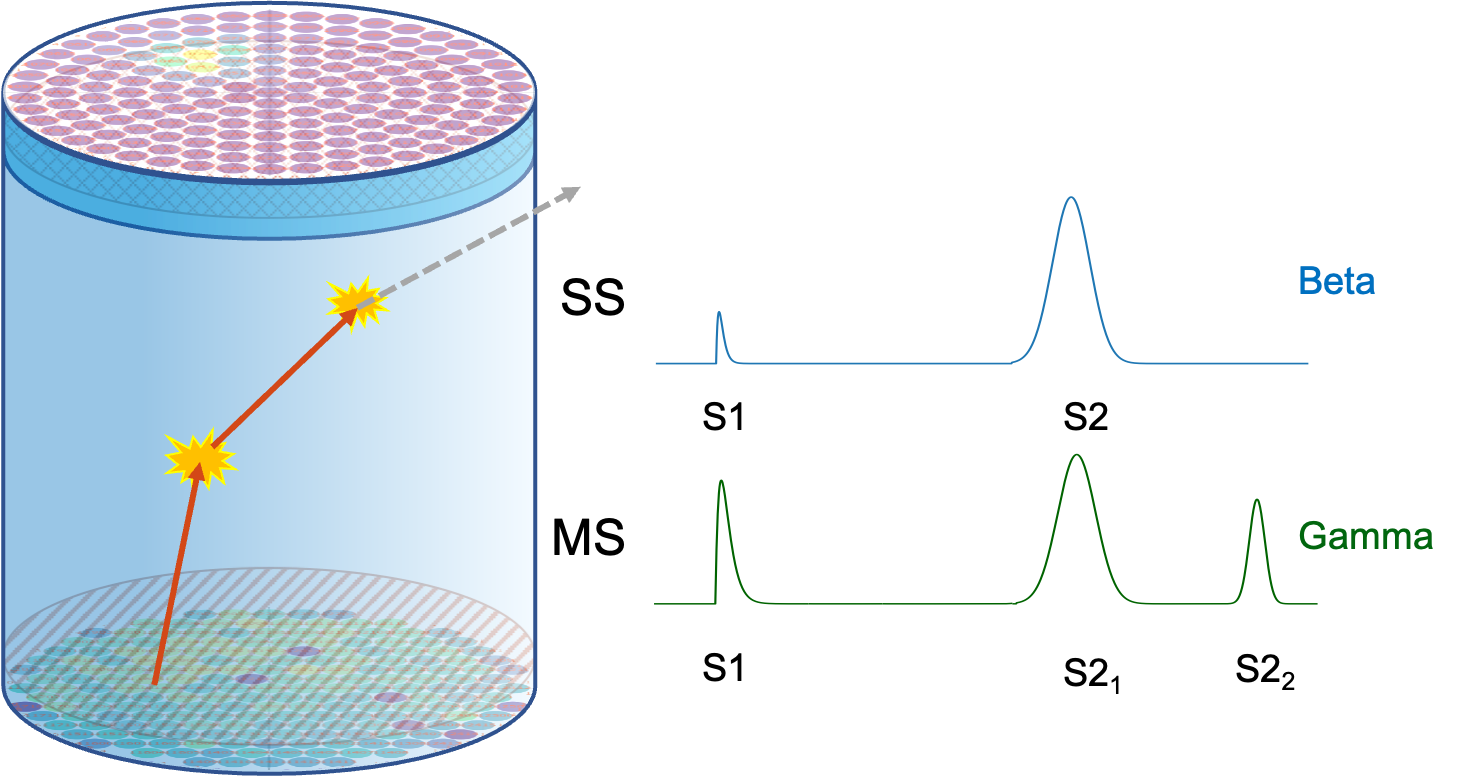}
  \caption{(Left) Schematic drawing of PandaX-4T TPC with the top and bottom PMT arrays. (Right) Illustration of typical single-site (SS) and Multi-site (MS) signals in the TPC.}
  \label{fig:detector}
\end{figure}

The PandaX-4T experiment is located in the B2 hall of the China Jinping Underground Laboratory (CJPL)~\cite{Cheng:2017usi,Li:2014rca}.
The large cylindrical xenon TPC has a sensitive volume of 3.7 tonnes of LXe.
The sensitive volume is enclosed by an electric field cage, as well as top and bottom electrodes, all of which work together to create a uniform electric field.
The field cage is covered by 24 highly reflective polytetrafluoroethylene (PTFE) wall panels, designed to direct photons to the top and bottom PMT arrays as effectively as possible.
169 and 199 Hamamatsu R11410-23 3-inch PMTs were instrumented for the top and bottom PMT arrays, respectively, shown in Fig~\ref{fig:detector}.
A detailed introduction of the PandaX-4T detector can be found in~\cite{PandaX-4T:2021bab}.

In PandaX-4T, the top and bottom PMT arrays record the prompt scintillation signals (S1) and the delayed electroluminescence signals (S2) produced by energy deposition in the active volume of the TPC.
The S1 signal originates at the event vertex and scintillates in all directions.
Considering solid angles and reflection on the PTFE panels, S1 results in relatively uniform hit patterns on both the top and bottom arrays.
The S2 signal scales with the number of ionization electrons.
These electrons drift towards the surface of the LXe under the influence of the electric field and are subsequently extracted into the region of gaseous xenon (GXe).
In this region, electrons undergo proportional amplification, accompanied by the generation of S2 photons by electroluminescence.
Situated approximately 6 cm below the top PMT array, the S2 signal exhibits a concentrated hit pattern on the top array (S2T).
The brightest top PMT above the electroluminescence track can potentially receive up to 30\% of the total S2T charge.
The concentrated charge pattern is essential for the horizontal position reconstruction of an event.
The bottom PMT array receives more uniform light, which is used for energy reconstruction.
The time delay between the S1 and S2 signals determines the depth of interaction along the vertical direction.

The energy and three-dimensional (3D) positional information obtained from the PMT arrays directly determine the physics reach of LXe detectors.
The reconstructed energy defines the region of interest for dark matter and neutrino physics.
The relative sizes of S1 and S2 distinguish between electron-recoil and nuclear-recoil events.
The 3D position determines the boundaries of fiducial volumes (FV) for various analyses.
The number of S1 and S2 signals in an event is essential to distinguish between single-site (SS) and multiple-site (MS) events.

The number of photoelectrons (PEs) collected by a PMT ranges from one to $3\times10^5$ for an energy range from sub-keV to multiple MeV.
For a 2.6 MeV gamma emitted by the $^{232}$Th decay chain, the expected number of electroluminescence photons is over one million in PandaX-4T based on NEST model~\cite{Szydagis:2021hfh}.
The typical quantum efficiency of the PMT is approximately 30\% for the 175 nm wavelength of xenon scintillation light.
The S2 signal produces approximately one million PEs in the top array and 350,000 PEs in the bottom array.
Therefore, the brightest PMT on the top array may receive $3\times10^5$ PEs or more.
The number of PEs cannot be reduced by lowering the PMT gain since a high gain is required to maintain a high detection efficiency for a single PE (SPE)~\cite{Yang:2021hnn}.
The typical PMT gain is $5.5\times 10^6$ $\rm{e}^-$ with a high voltage of 1500 V between the photocathode and the anode.

\begin{figure}[tbp]
  \centering
  \includegraphics[width=\textwidth]{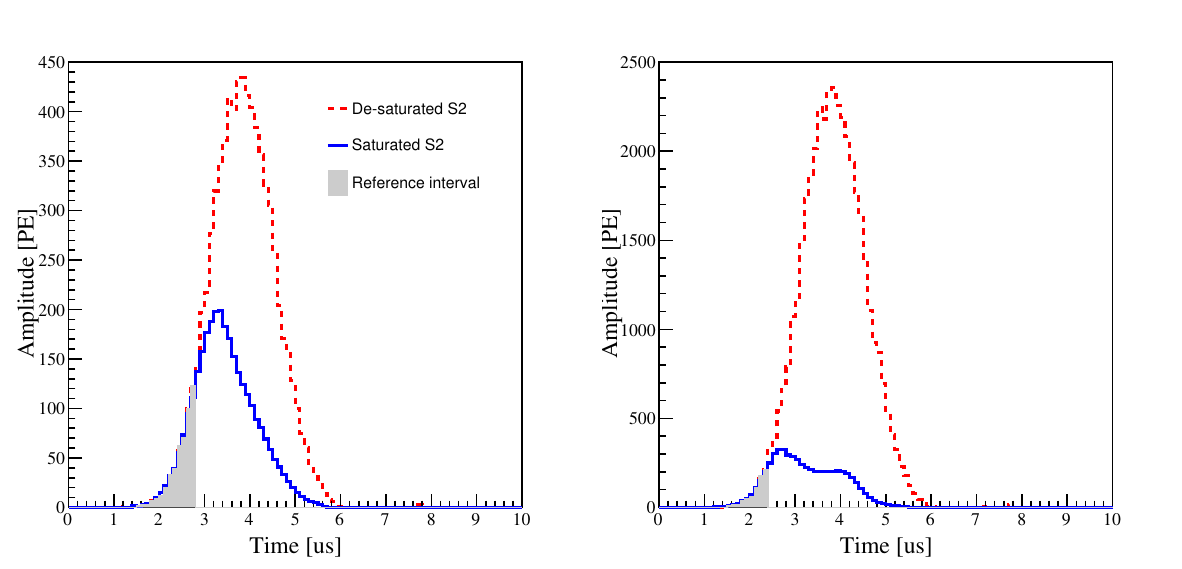}
  \caption{Illustration of saturated waveforms and the de-saturation algorithm. The blue waveform in the left (right) figure has an observed charge of approximately 1800 (7800) PEs. After the de-saturation, the corrected charge is approximately 4400 (56000) PEs, as demonstrated by the red dashed waveform.}
  \label{fig:satur stage}
\end{figure}

\subsection{Saturation effect of high energy events}
In the PandaX-4T commissioning run, it has been observed that the initial PMT base design leads to significant saturation of the individual PMTs when the signal exceeds 1000 PEs.
The saturation is manifested by distorted waveforms, as illustrated in Fig.~\ref{fig:satur stage}.
The blue-saturated ones exhibit distinct distortion features compared to the dashed reference waveform.
Both saturated waveforms peak at an earlier time than the reference one.
The waveform shows a distinctive double-peak feature for the blue waveform in the right figure with an even more significant light exposure.
However, both waveforms maintain the same rising edge shape as the reference.

A de-saturation algorithm is developed to mitigate PMT waveform distortion.
The summed PE value of a distorted waveform no longer reflects the amount of photons collected.
The deviation increases as the intensity of the light increases.
To address this, we first define a reference waveform by aggregating non-saturated waveforms of the signal of the same event.
The PMTs from the top and bottom array with charges between 50-900~PE are selected as reference PMTs.The reference waveform is calculated for each signal since it contains the timing information of photons arriving at the PMT arrays.
The upper threshold of 900~PE for channel waveform selection is optimized by scanning various charge thresholds to minimize the energy resolution at 2.6~MeV from Th-232 calibration events.
For a typical 2.6~MeV gamma event, multiple (>10) non-saturated waveforms can always be identified.
The summation of these waveforms is intended to minimize fluctuations arising from the rising edges of the individual PMT waveforms.
The summation also averages out the minor timing profile differences of PMTs at different positions.
The reference waveform is then scaled to align the rising edge with the saturated waveforms, as shown in Fig.~\ref{fig:satur stage}.
The optimal scale factor is achieved by minimizing the chi-square values between the two waveforms.
The de-saturated charge is the number of PEs of the scaled reference waveform.

The de-saturation algorithm presumes that the rising edge of a waveform remains unaffected even if the later part of the waveform is distorted.
The presumption is validated by the improved detector position and energy reconstruction improvement in the PandaX-4T data~\cite{silin2022DBD}. 
However, this method has a large uncertainty due to the fluctuation of the rising edge of the saturated waveform.
The bench test data further confirms it, as we will describe later.

De-saturation is unnecessary for S1 signals.
For a 2.6 MeV gamma event, the typical S1 charge on the order of $10^4$ PEs is more or less evenly distributed on all PMTs of the top and bottom arrays.
Therefore, the charge of each PMT is much smaller than the 900-PE threshold.

\subsection{Suppression effect of multi-scattering events}

A large signal may also distort subsequent signals, which we call suppression.
The cause of PMT waveform distortion is not intrinsic to the PMT itself but from voltage instabilities between dynode stages within the PMT due to the high voltage divider base.
The suppression of nearby signals was observed in the PandaX-4T commissioning.
As shown in Fig.~\ref{fig:detector}, Compton scattering of $\gamma$ rays in the detector creates multiple energy deposition vertices in the detector, which results in a multi-scattering (MS) event with multiple S2 signals.
For the gamma Compton scattering, the prompt S1s are overlapped, but the S2 waveform may have separate peaks depending on the vertical position.
We identify peaks on the S2 waveform and label SS and MS events by a vertical position-based algorithm~\cite{silin2022DBD}.
During an MS event, an individual PMT receiving a significant number of photons from one vertex may get saturated and lose the response to the subsequent S2 signal.

The extent of this suppression effect on the second S2 (S2$_2$) is intrinsically related to the magnitude of the charge of the first S2 (S2$_1$) and the time interval between the two S2 signals $\Delta$T.
A large S2$_1$ signal and a shorter time interval between S2$_1$ and S2$_2$ bring more prominent suppression to the second S2 signal.
In an extreme case, a PMT may experience complete suppression of the second S2 signals as the PMT's response is fully compromised for a particular duration due to an extremely large S2$_1$.
The compromised response may last up to a few tens of microseconds.
The phenomenon is not different from the saturation effect shown in the right plot of Fig.~\ref{fig:satur stage}, where the saturated waveform approaches zero between the 5 and 6 $\mu$s mark. 
At the same time, there are still photons arriving according to the reference waveform.

\begin{figure}[tbp]
  \centering
  \includegraphics[width=0.7\textwidth]{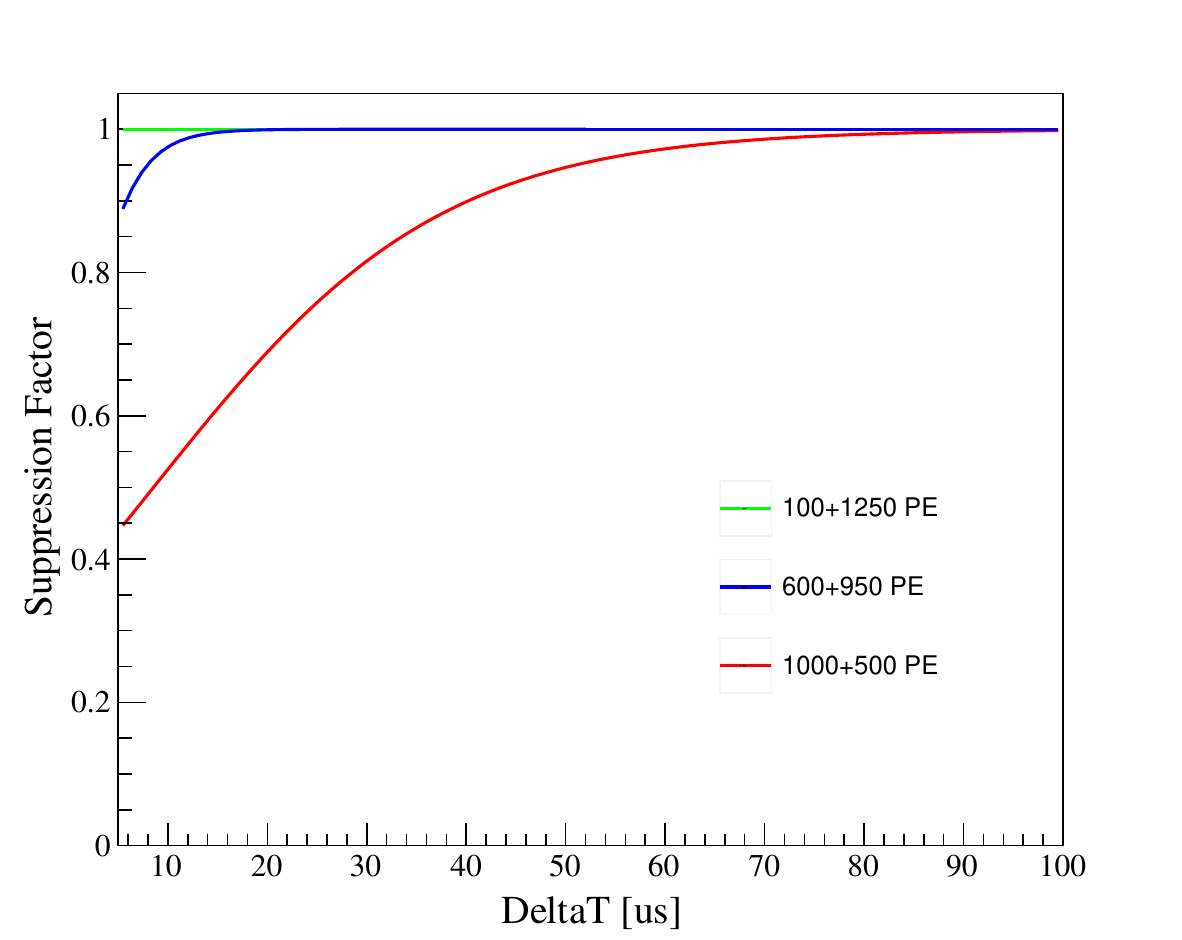}
  \caption{The suppression factor of $S2_2$ by the $S2_1$ as a function of the time interval $\Delta$T between the two signals. Different lines represent different charges of $S2_1$.}
  \label{fig:suppression factor}
\end{figure}

We quantify the suppression effect with double-site events of 2.6 MeV gamma in the PandaX-4T commissioning data. The suppression factor is defined as the charge ratio of observed S2$_2$ and expected S2$_2$ 100$\mu$s away from S2$_1$, where the suppression effect becomes unnoticeable with this large time gap. 
In Fig.~\ref{fig:suppression factor}, PMTs with 100, 600, and 1000~PEs in the S2$_1$ signals are selected, correspondingly the actually S2$_2$ signals are 1250, 950 and 500~PE. The suppression factor of the second site as a function of the time interval between the second and the first signals is shown.
As one can see, even for a moderately large S2$_1$ waveform of 600 PE, the second waveform may be suppressed to approximately 90\% when the time interval is small.
For a large S2$_1$ waveform of 1000 PE, the second waveform can be only half the size without suppression.
Considering the typical drift velocity of 1.4 mm/$\mu$s during the PandaX-4T commissioning runs, the "safe" vertical distance should be more than 14 cm for a preceding 1000-PE waveform.

The suppression effect has a considerable impact on the PandaX-4T analysis.
The sequential multi-Compton scattering sites are usually a few centimeters away for MeV gamma events.
Therefore, multiple sites are not separated enough vertically to avoid suppression.
Horizontally, the sites are so close that S2$_2$ is suppressed for PMTs with significant charge information.
Often, the PMT charge pattern of S2$_2$ is deformed, and proper position reconstruction is impossible.
The energy reconstruction suffers as well.
Therefore, an improved PMT base is required for better detector performance in the MeV energy range.

\section{Improved PMT base}
\label{sec:new design base}

\begin{figure}[tbp]

\centering 
\subfigure[The original PMT base with three capacitors]{
\label{fig:base original design}
\centering
\includegraphics[width=\textwidth]{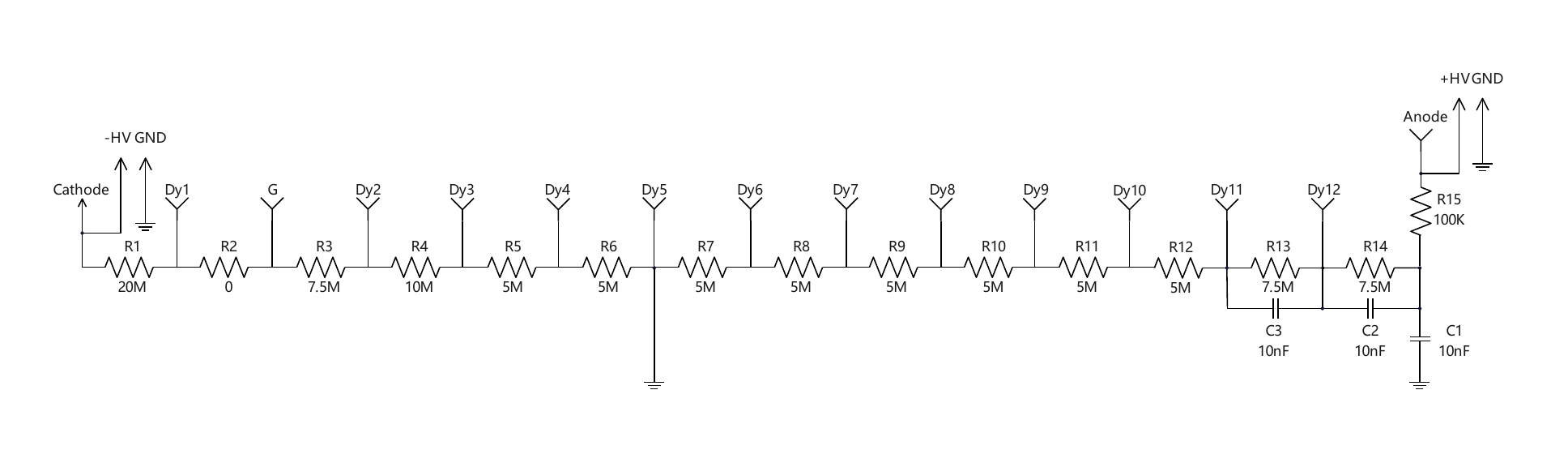}
}
\subfigure[The improved PMT base with six capacitors]{
\label{fig:base v2}
\centering
\includegraphics[width=\textwidth]{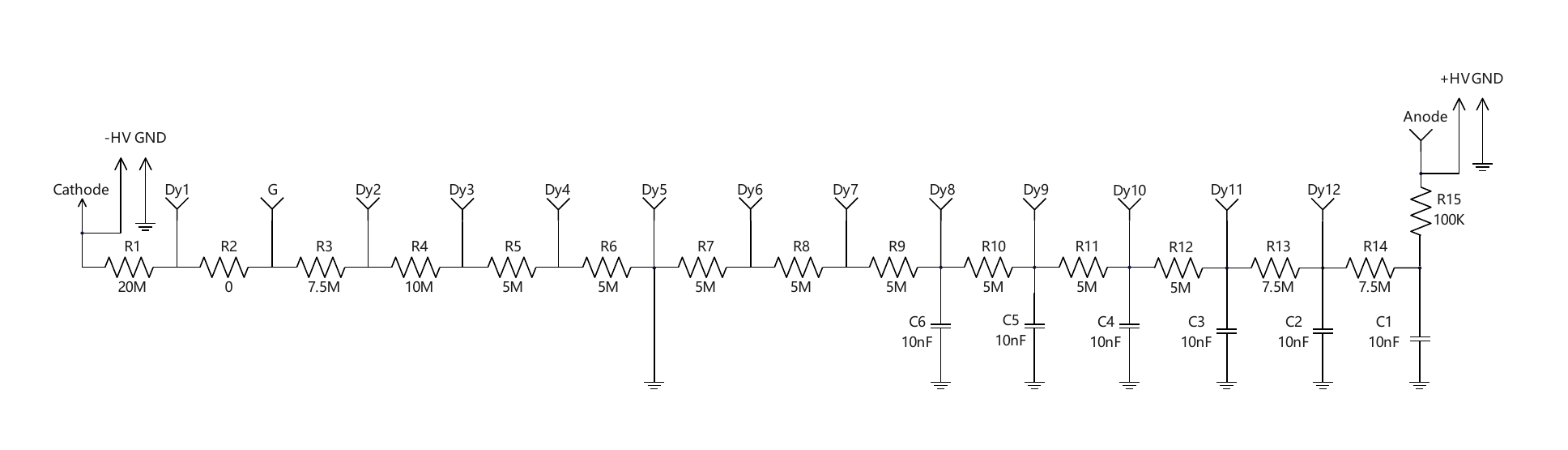}
}
\caption{The schematics of the original (a) and improved (b) PMT bases. }
\label{fig:base figure}
\end{figure}

\begin{figure}[tbp]
  \centering
  \includegraphics[width=\textwidth]{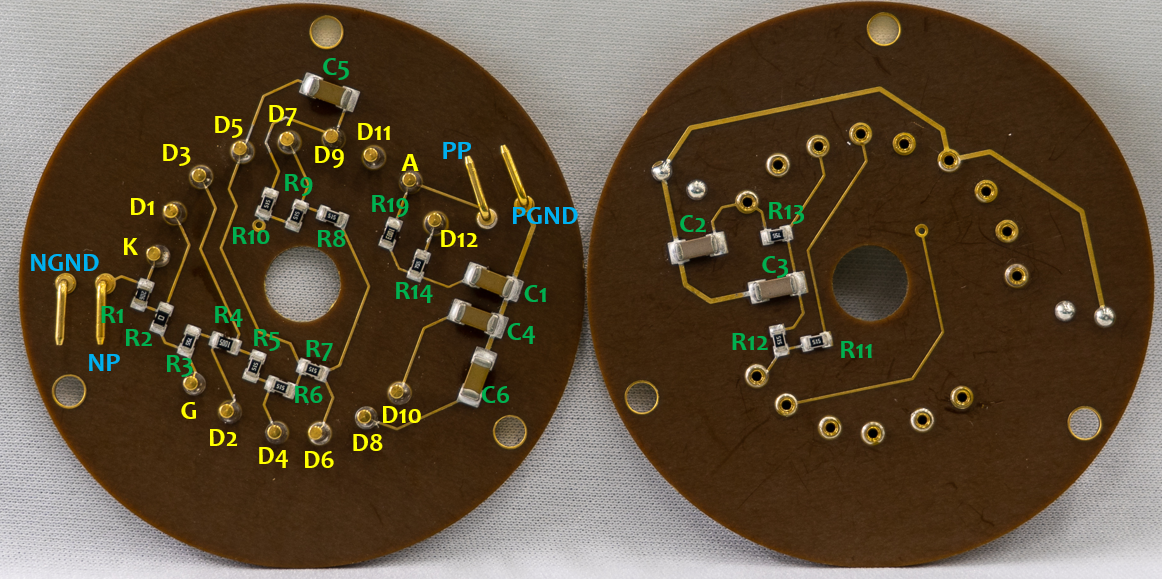}
  \caption{Photo of the front and back of the new high voltage divider base for Hamamatsu R11410-023 PMT. The dynode pins are marked in yellow, the resistors and capacitors are in green, and the high voltage and ground pins are in blue.}
  \label{fig:base photo}
  \end{figure}

The original and improved PMT base designs are shown in Fig.~\ref{fig:base figure}, and a photo of the improved version is in Fig.~\ref{fig:base photo}. 
The PMT base is a printed circuit board (PCB) made of two-layer Kapton using hot-pressing technology.
A split positive and negative high voltage scheme was used with the ground on the fifth dynode (DY5) to reduce the relative potential to the ground.
Such a design can help reduce the high voltage tolerance requirement of capacitors and the risk of discharge on the base, cables,  and feedthrough pins. 
Two separate coaxial cables are plugged into gold-plated HV pins to supply the PMT with high voltage and read out the anode signal. 

The design of PandaX-4T PMT bases emphasizes low radioactivity requirements and tries to limit the number of capacitors, which is usually more radioactive than other components.
Besides the capacitor C1 to ground, the original base used two decoupling capacitors at the last stages, which maintains voltage between two dynodes when the electrical pulse is being read out and improves the linearity of the PMT.
In the improved design, we add three more decoupling capacitors from DY8 to DY10, and each capacitor is connected to the ground directly, as shown in Fig.~\ref{fig:base v2}. 
As we will show later, the additional capacitors improve the dynamic range of the anode significantly. 
A new model of Knowles capacitor is found with only half the radioactivity of the current 4T capacitor, based on the assay results from two high purity germanium (HPGe) counting stations in CJPL. The radioactivity of the new base is at the same level as the initial one~\cite{PandaX-4T:2021lbm}.

The design and fabrication of PMT bases comply with cryogenic requirements. 
We conducted three thermal cycles for every PMT base from -100$\rm{^o}$C to room temperature.
We then measured every resistor and capacitor to ensure the connectivity. 
A high voltage test up to 1000 V was performed as the final quality control.

\section{Bench test of the new base}
\label{sec:bench test}

\subsection{Bench test setup}
\begin{figure}[tbp]
  \centering
  \includegraphics[width=\textwidth]{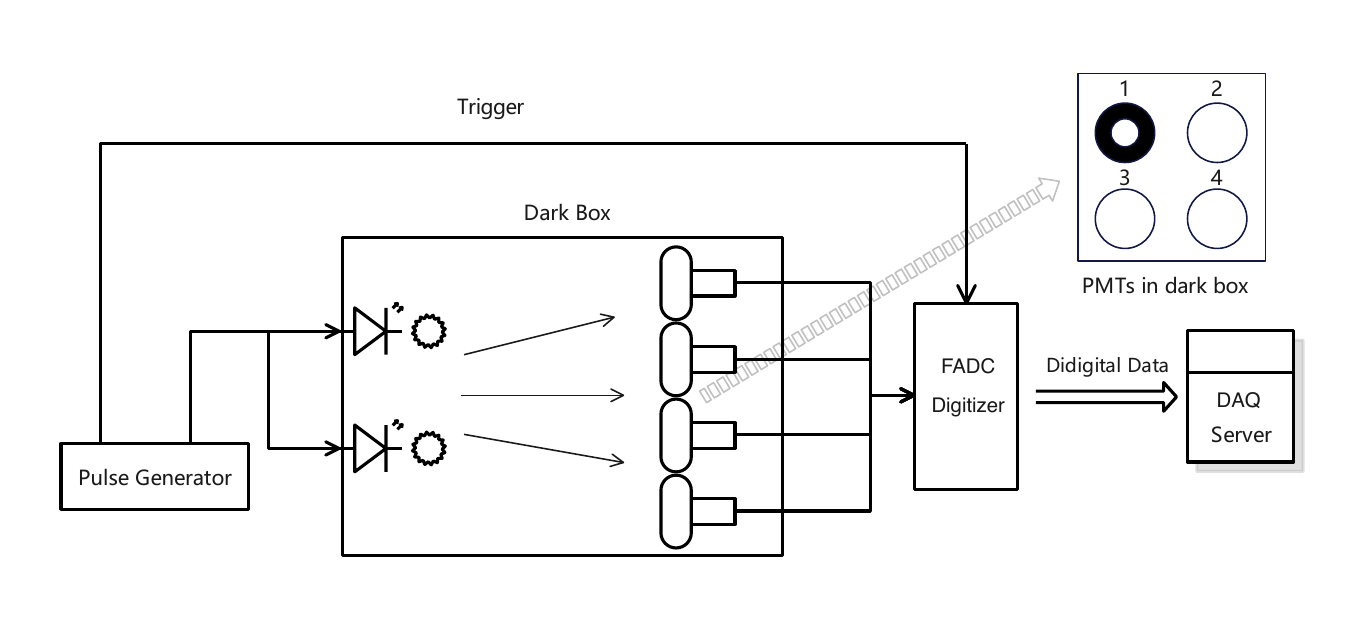}
  \caption{The PMT bench test setup with main components marked. The four PMTs on the top right are equipped with (1) an improved base and a screen to control 10\% of light exposure, (2) an improved base, (3) an original base, and (4) an original base. PMT 1 is used for monitoring.}
  \label{fig:offline app}
\end{figure}

To confirm the saturation and suppression symptoms of the original PMT base, validate the de-saturation algorithm independently, and benchmark the performance of the improved base, we set up a dedicated PMT bench test as shown in Fig.~\ref{fig:offline app}. 
The system consists of four R11410-23 PMTs equipped with different bases, an arbitrary pulse generator (RIGOL DG4162), two blue light LEDs, and a data acquisition (DAQ) system. 

The four PMTs are instrumented as follows.
Two PMTs are equipped with original bases and one with an improved base for testing.
One last PMT used as a monitor PMT is covered by a black screen with a central aperture in the center.
The aperture controls the incident light intensity to approximately one-tenth of that received by the other PMTs.
The DG4162 dual-channel pulse generator is used to drive the LEDs.
The two channels can be synchronized phase-dependently to generate a correlated light output timing profile to mimic an S2-like waveform or a plain square waveform.
A diffusion ball attached to each LED established light output as isotropic as possible.
The DAQ system works under external trigger mode with a trigger synchronized from the pulse generator.
The DAQ records the PMT waveforms with a 500 MHz sampling rate and a sampling window of 7000 samples. 
The PMT charge is calculated by summing the entire waveform in the sampling window. 

The light output is tuned for different purposes.
To avoid  PMT damage under excessive exposure, the average anode current per second is limited to 0.1~mA~\cite{Hamamatsu-hand-book} during the test.
For saturation measurement, the LEDs emit light that could generate PMT waveforms of a few hundred PEs to hundreds of thousands of PEs.
For the suppression test, two pulses with time interval $\Delta$T from several $\mu$s to 100 $\mu$s are generated synchronously. 

The performance of the bases is tested under different PMT gain and pulse shapes. 
The PMT gain is modified from 3 $\times 10^6\rm{e}^-/\rm{SPE}$ to 5 $\times 10^6\rm{e}^-/\rm{SPE}$. 
The consistency of the gain before and after the bench test is within 1\%.  
To ensure the reliability of the conclusions, we conducted multiple rounds of tests by pairing four PMTs and four bases interchangeably.

\subsection{Extending of linearity range}

The PMT saturation effect is confirmed by comparing the response of test PMTs and monitor PMT.
Since the monitor PMT has an improved base and a screen to block approximately 90\% of the light, it remains in the linear response region in our test.
We also use the anode charge in the number of electrons instead of photoelectrons as the saturation benchmark since the anode charge also incorporates the influence of PMT gain.
The PMT gain is measured by the integral of the SPE waveform. 
To ensure the reliability of the gain, we perform PMT gain calibration before and after each test. 

\begin{figure}[tbp]
  \centering
  \includegraphics[scale=0.4]{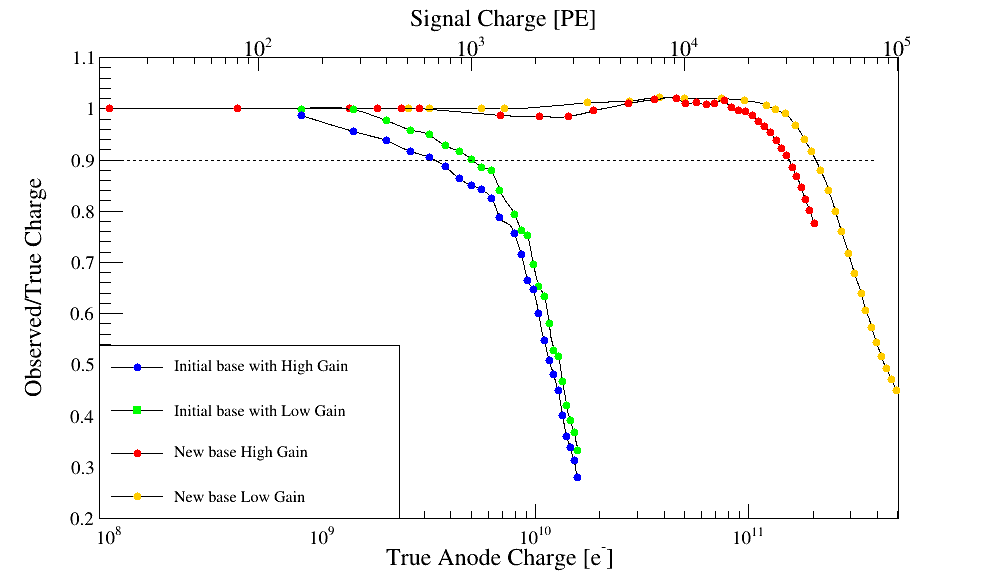}
  \caption{PMT linearity as a function of the anode current. The blue and green dots represent the test results with the original bases, while the red and yellow dots represent the test results with the improved bases. Under high gains, the corresponding numbers of photoelectrons for serious saturation (defined as a suppression factor of 90\%) are approximately 1000 PEs and 40000 PEs for the original and improved bases, respectively.}
  \label{fig:max_charge_per_pmt}
\end{figure}

The saturation is correlated with the total anode charge and not influenced by the waveform shape, as shown in Fig.~\ref{fig:max_charge_per_pmt}.
The horizontal axis represents the number of electrons collected by the PMT anode in response to an LED light signal.
The vertical axis represents the ratio of the measured anode charge to the expected anode charge, which is scaled from monitor PMT anode charges.
The ratio represents the linearity of PMT response, and any deviation from the unit means non-linearity. 
If we define the threshold at 90\%, PMTs with the original bases show severe saturation at anode charge of $3.8\times 10^9$ and $5.0\times 10^9$ electrons for high and low PMT gains, respectively.
For PMTs with improved bases, the threshold corresponds to $1.5\times 10^{11}$ and $2.1\times 10^{11}$ electrons. 
If we convert the anode charge to the number of PEs, the linear response range of the PMT has been extended to approximately 40000~PE with the improved base.

It can be observed that reducing the PMT gain helps improve the linearity range in terms of the number of photoelectrons.
However, considering the requirements for the SPE trigger rate in the PandaX-4T DAQ for low-energy physics analyses, we generally prefer the gain to be 5$\times 10^6\rm{e}^-/SPE$ or higher.

\begin{figure}[tbp]
  \centering
  \includegraphics[scale=0.4]{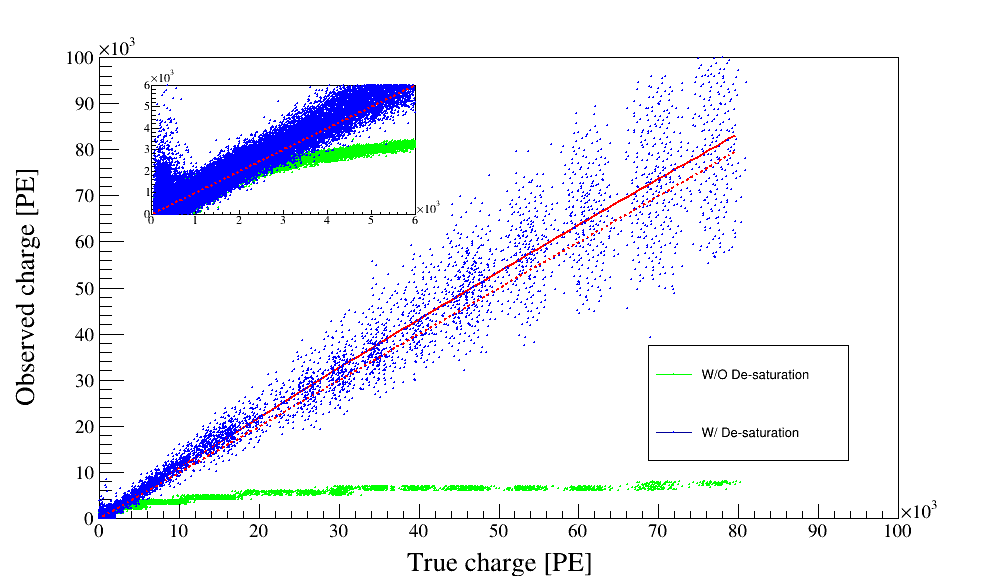}
  \caption{Observed charge with and without de-saturation. The red line is the mean of de-sturation charge and the red dashed line is the y=x reference line. The inset highlights points with PMT true charge of fewer than 6000 PEs.}
  \label{fig:de-saturation}
\end{figure}

The de-saturation algorithm demonstrated in Fig.~\ref{fig:satur stage} 
is validated with our bench test.
Fig.~\ref{fig:de-saturation} shows the observed charge and corrected charge with de-saturation versus the true charge, which is again calculated from the monitor PMT. S2-like waveform with a width of several $\mu$s is used in the test. When the true charge is a few hundred, the Gauss-like shape of the waveform is not valid, resulting in the unexpected tail at the beginning.
When the number of true charges is more than $10^4$ PEs for PMTs with the original base, the measured charge deviates significantly from expectation.
The de-saturation correction algorithm can rectify this over an extensive dynamic range.
Nevertheless, uncertainties increase with higher charges due to the severity of saturation, where the saturated waveform loses sensitivity to reflect the true charge accurately. Although de-saturation can alleviate the problem, an improved base is a more effective solution.

\begin{figure}[tbp]
  \centering
  \includegraphics[scale=0.72]{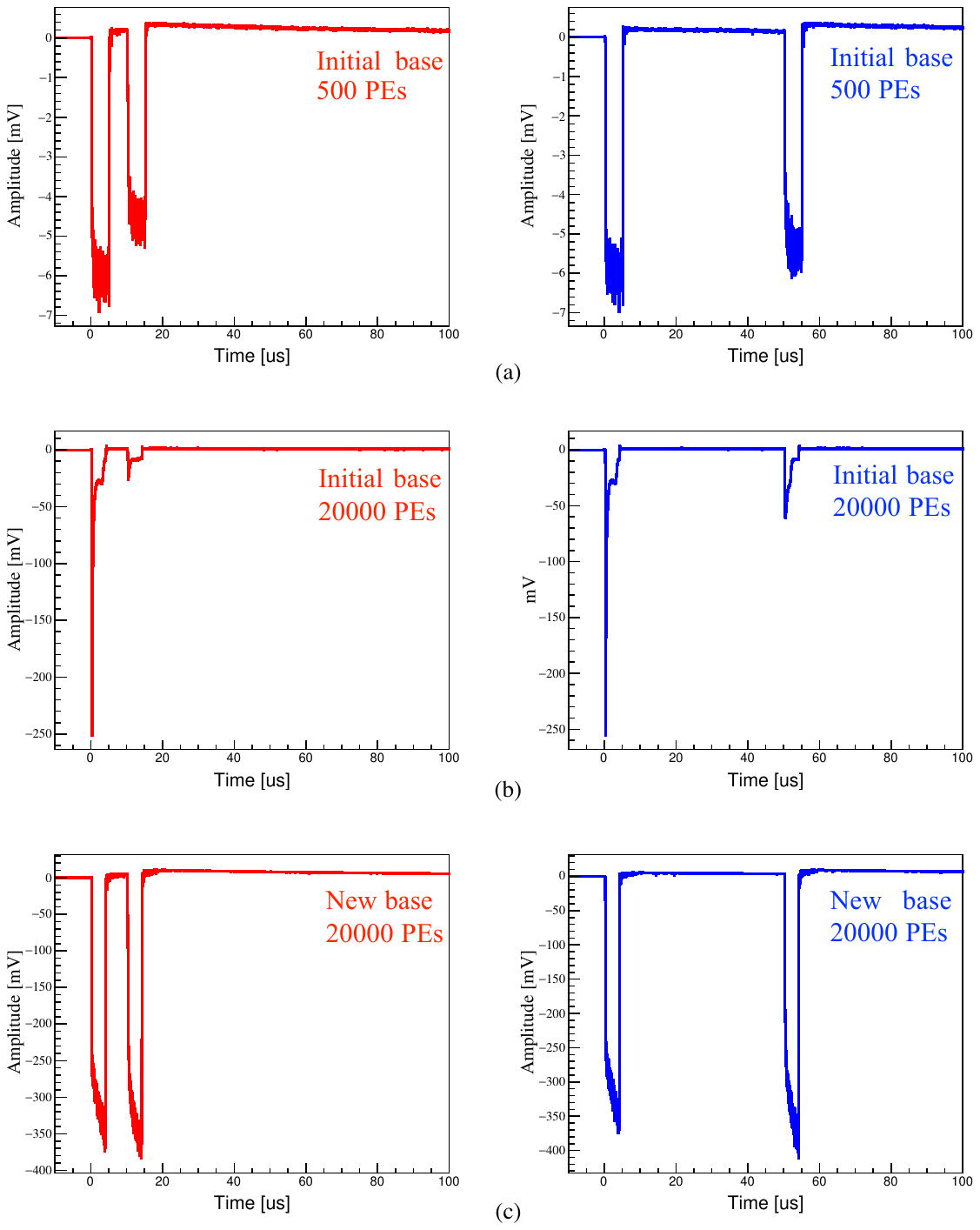}
  \caption{Typical example waveforms for illustrating the suppression effect of the first signal on the second signal. For each figure, the input light intensities for the two signals are approximately the same; therefore, we expect two signals of the same size. (a) Original bases with an expected signal charge of 500 PEs, signal intervals of 10 $\mu$s (left) and 50 $\mu$s (right). (b) Original bases with an expected signal charge of over 20000 PEs, signal intervals of 10 $\mu$s (left) and 50 $\mu$s (right). (c) Improved bases with an expected signal charge of over 20000 PEs, signal intervals of 10 $\mu$s (left) and 50 $\mu$s (right).}
  \label{fig:sup waveform}
\end{figure}

The PMT suppression is verified qualitatively with our bench test.
In the tests, the light intensity and timing profile of the two LEDs were fixed, and the time interval between the emissions of the two LED lights was adjusted to mimic double-site events in the PandaX-4T detector. 
The time intervals range from 7 $\mu$s to 100 $\mu$s.
Fig.~\ref{fig:sup waveform} (a) and (b) illustrate the suppression effect to the second waveform for PMTs with the original bases.
In Fig.~\ref{fig:sup waveform} (a), a relatively small charge of 500~PE in the first waveform still causes noticeable suppression to the second waveform right after it.
When the time interval increases to 47~$\mu$s, the suppression is less serious than the first case.
If we increase the first PMT waveform to have more than 20000 PEs, the suppression effect is much more pronounced, even if the time interval increases, as demonstrated in Fig.~\ref{fig:sup waveform} (b).
One can also notice the distortion for the first and second waveforms in Fig.~\ref{fig:sup waveform} (b).
However, with the use of the improved PMT base, the suppression effect is no longer apparent (Fig.~\ref{fig:sup waveform} (c)). 
The suppression effect is no longer visible from our qualitative tests with a summed charge below 40000~PE.

\section{Summary}

We report the PMT saturation and suppression issues encountered during the commissioning of the PandaX-4T.
Large PMT signal with MeV scale events causes a temporary voltage instability at the later dynode, which leads to fluctuations in the multiplication effect of the PMT and induces the non-linear response of the PMT.
We have developed a de-saturation algorithm to mitigate this issue in PandaX-4T data. 
The de-saturation algorithm has been successfully utilized in data analysis and validated with our bench test described in this paper.

To address these issues further, we have redesigned the voltage divider base to include six de-saturation capacitors while ensuring that the overall radioactivity of the base remains unchanged. 
PMTs with the new base achieve a linearity range of up to 40000 PEs.

The extension of the linearity response of PMT is crucial for the analysis of MeV-scale events in PandaX-4T.
The newly improved PMT bases have been installed in PandaX-4T.
We expect the energy response in the MeV range will be significantly improved, enhancing the detector's sensitivity in the search for astrophysical neutrinos, neutrinoless double-beta decay, and other physics topics in this energy range.

\section{Acknowledgement}
This project is supported in part by grants from the National Science Foundation of China (Nos. 12105052,
12090060, 12005131, 11905128, 11925502), and by the Office of Science and Technology, Shanghai Municipal Government (grant Nos. 18JC1410200, 22JC1410100). 
We thank the support from the Double First Class Plan of Shanghai Jiao Tong University. 
We also thank the sponsorship from the Chinese Academy of Sciences Center for Excellence in Particle Physics (CCEPP), Hongwen Foundation in Hong Kong, and Tencent Foundation in China.

\bibliography{output.bbl}

\begin{thebibliography}{10}
\expandafter\ifx\csname url\endcsname\relax
  \def\url#1{\texttt{#1}}\fi
\expandafter\ifx\csname urlprefix\endcsname\relax\def\urlprefix{URL }\fi
\expandafter\ifx\csname href\endcsname\relax
  \def\href#1#2{#2} \def\path#1{#1}\fi

\bibitem{Aalbers:2022dzr}
J.~Aalbers, et~al., {A next-generation liquid xenon observatory for dark matter
  and neutrino physics}, J. Phys. G 50~(1) (2023) 013001.
\newblock \href {http://arxiv.org/abs/2203.02309} {\path{arXiv:2203.02309}},
  \href {https://doi.org/10.1088/1361-6471/ac841a}
  {\path{doi:10.1088/1361-6471/ac841a}}.

\bibitem{DARWIN:2016hyl}
J.~Aalbers, et~al., {DARWIN: towards the ultimate dark matter detector}, JCAP
  11 (2016) 017.
\newblock \href {http://arxiv.org/abs/1606.07001} {\path{arXiv:1606.07001}},
  \href {https://doi.org/10.1088/1475-7516/2016/11/017}
  {\path{doi:10.1088/1475-7516/2016/11/017}}.

\bibitem{PandaX-4T:2021bab}
Y.~Meng, et~al., {Dark Matter Search Results from the PandaX-4T Commissioning
  Run}, Phys. Rev. Lett. 127~(26) (2021) 261802.
\newblock \href {http://arxiv.org/abs/2107.13438} {\path{arXiv:2107.13438}},
  \href {https://doi.org/10.1103/PhysRevLett.127.261802}
  {\path{doi:10.1103/PhysRevLett.127.261802}}.

\bibitem{XENON:2020kmp}
E.~Aprile, et~al., {Projected WIMP sensitivity of the XENONnT dark matter
  experiment}, JCAP 11 (2020) 031.
\newblock \href {http://arxiv.org/abs/2007.08796} {\path{arXiv:2007.08796}},
  \href {https://doi.org/10.1088/1475-7516/2020/11/031}
  {\path{doi:10.1088/1475-7516/2020/11/031}}.

\bibitem{LZ:2021xov}
D.~S. Akerib, et~al., {Projected sensitivities of the LUX-ZEPLIN experiment to
  new physics via low-energy electron recoils}, Phys. Rev. D 104~(9) (2021)
  092009.
\newblock \href {http://arxiv.org/abs/2102.11740} {\path{arXiv:2102.11740}},
  \href {https://doi.org/10.1103/PhysRevD.104.092009}
  {\path{doi:10.1103/PhysRevD.104.092009}}.

\bibitem{silin2022DBD}
L.~Si, et~al., {Determination of Double Beta Decay Half-Life of 136Xe with the
  PandaX-4T Natural Xenon Detector}, Research 2022 (2022) 9798721.
\newblock \href {http://arxiv.org/abs/2205.12809} {\path{arXiv:2205.12809}},
  \href {https://doi.org/10.34133/2022/9798721}
  {\path{doi:10.34133/2022/9798721}}.

\bibitem{2022XenonDBD}
E.~Aprile, et~al., {Double-Weak Decays of $^{124}$Xe and $^{136}$Xe in the
  XENON1T and XENONnT Experiments}, Phys. Rev. C 106~(2) (2022) 024328.
\newblock \href {http://arxiv.org/abs/2205.04158} {\path{arXiv:2205.04158}},
  \href {https://doi.org/10.1103/PhysRevC.106.024328}
  {\path{doi:10.1103/PhysRevC.106.024328}}.

\bibitem{LZ:2019qdm}
D.~S. Akerib, et~al., {Projected sensitivity of the LUX-ZEPLIN experiment to
  the $0\nu\beta\beta$ decay of $^{136}Xe$}, Phys. Rev. C 102~(1) (2020)
  014602.
\newblock \href {http://arxiv.org/abs/1912.04248} {\path{arXiv:1912.04248}},
  \href {https://doi.org/10.1103/PhysRevC.102.014602}
  {\path{doi:10.1103/PhysRevC.102.014602}}.

\bibitem{PandaX-II:2019euf}
K.~Ni, et~al., {Searching for neutrino-less double beta decay of $^{136}$Xe
  with PandaX-II liquid xenon detector}, Chin. Phys. C 43~(11) (2019) 113001.
\newblock \href {http://arxiv.org/abs/1906.11457} {\path{arXiv:1906.11457}},
  \href {https://doi.org/10.1088/1674-1137/43/11/113001}
  {\path{doi:10.1088/1674-1137/43/11/113001}}.

\bibitem{XENON:2019dti}
E.~Aprile, et~al., {Observation of two-neutrino double electron capture in
  $^{124}$Xe with XENON1T}, Nature 568~(7753) (2019) 532--535.
\newblock \href {http://arxiv.org/abs/1904.11002} {\path{arXiv:1904.11002}},
  \href {https://doi.org/10.1038/s41586-019-1124-4}
  {\path{doi:10.1038/s41586-019-1124-4}}.

\bibitem{4TB8}
W.~Ma, et~al., {Search for Solar B8 Neutrinos in the PandaX-4T Experiment Using
  Neutrino-Nucleus Coherent Scattering}, Phys. Rev. Lett. 130~(2) (2023)
  021802.
\newblock \href {http://arxiv.org/abs/2207.04883} {\path{arXiv:2207.04883}},
  \href {https://doi.org/10.1103/PhysRevLett.130.021802}
  {\path{doi:10.1103/PhysRevLett.130.021802}}.

\bibitem{XENON:2020gfr}
E.~Aprile, et~al., {Search for Coherent Elastic Scattering of Solar $^8$B
  Neutrinos in the XENON1T Dark Matter Experiment}, Phys. Rev. Lett. 126 (2021)
  091301.
\newblock \href {http://arxiv.org/abs/2012.02846} {\path{arXiv:2012.02846}},
  \href {https://doi.org/10.1103/PhysRevLett.126.091301}
  {\path{doi:10.1103/PhysRevLett.126.091301}}.

\bibitem{Baudis:2013xva}
L.~Baudis, A.~Behrens, A.~Ferella, A.~Kish, T.~Marrodan~Undagoitia, D.~Mayani,
  M.~Schumann, {Performance of the Hamamatsu R11410 Photomultiplier Tube in
  cryogenic Xenon Environments}, JINST 8 (2013) P04026.
\newblock \href {http://arxiv.org/abs/1303.0226} {\path{arXiv:1303.0226}},
  \href {https://doi.org/10.1088/1748-0221/8/04/P04026}
  {\path{doi:10.1088/1748-0221/8/04/P04026}}.

\bibitem{Zheng:2020kfp}
Q.~Zheng, et~al., {An improved design of the readout base board of the
  photomultiplier tube for future PandaX dark matter experiments}, JINST
  15~(12) (2020) T12006.
\newblock \href {http://arxiv.org/abs/2012.03202} {\path{arXiv:2012.03202}},
  \href {https://doi.org/10.1088/1748-0221/15/12/T12006}
  {\path{doi:10.1088/1748-0221/15/12/T12006}}.

\bibitem{XENON:2020iwh}
E.~Aprile, et~al., {Energy resolution and linearity of XENON1T in the MeV
  energy range}, Eur. Phys. J. C 80~(8) (2020) 785.
\newblock \href {http://arxiv.org/abs/2003.03825} {\path{arXiv:2003.03825}},
  \href {https://doi.org/10.1140/epjc/s10052-020-8284-0}
  {\path{doi:10.1140/epjc/s10052-020-8284-0}}.

\bibitem{Cheng:2017usi}
J.-P. Cheng, et~al., {The China Jinping Underground Laboratory and its Early
  Science}, Ann. Rev. Nucl. Part. Sci. 67 (2017) 231--251.
\newblock \href {http://arxiv.org/abs/1801.00587} {\path{arXiv:1801.00587}},
  \href {https://doi.org/10.1146/annurev-nucl-102115-044842}
  {\path{doi:10.1146/annurev-nucl-102115-044842}}.

\bibitem{Li:2014rca}
J.~Li, X.~Ji, W.~Haxton, J.~S.~Y. Wang, {The second-phase development of the
  China JinPing underground Laboratory}, Phys. Procedia 61 (2015) 576--585.
\newblock \href {http://arxiv.org/abs/1404.2651} {\path{arXiv:1404.2651}},
  \href {https://doi.org/10.1016/j.phpro.2014.12.055}
  {\path{doi:10.1016/j.phpro.2014.12.055}}.

\bibitem{Szydagis:2021hfh}
M.~Szydagis, et~al., {A Review of Basic Energy Reconstruction Techniques in
  Liquid Xenon and Argon Detectors for Dark Matter and Neutrino Physics Using
  NEST}, Instruments 5~(1) (2021) 13.
\newblock \href {http://arxiv.org/abs/2102.10209} {\path{arXiv:2102.10209}},
  \href {https://doi.org/10.3390/instruments5010013}
  {\path{doi:10.3390/instruments5010013}}.

\bibitem{Yang:2021hnn}
J.~Yang, et~al., {Readout electronics and data acquisition system of PandaX-4T
  experiment}, JINST 17~(02) (2022) T02004.
\newblock \href {http://arxiv.org/abs/2108.03433} {\path{arXiv:2108.03433}},
  \href {https://doi.org/10.1088/1748-0221/17/02/T02004}
  {\path{doi:10.1088/1748-0221/17/02/T02004}}.

\bibitem{PandaX-4T:2021lbm}
Z.~Qian, et~al., {Low radioactive material screening and background control for
  the PandaX-4T experiment}, JHEP 06 (2022) 147.
\newblock \href {http://arxiv.org/abs/2112.02892} {\path{arXiv:2112.02892}},
  \href {https://doi.org/10.1007/JHEP06(2022)147}
  {\path{doi:10.1007/JHEP06(2022)147}}.

\bibitem{Hamamatsu-hand-book}
{Photomultiplier tubes: Basics and Applications},
  \url{https://www.hamamatsu.com/content/dam/hamamatsu-photonics/sites/documents/99_SALES_LIBRARY/etd/PMT_handbook_v4E.pdf}.

\end{thebibliography}
\bibliographystyle{elsarticle-num}

\end{document}